\documentclass[a4paper,12pt,leqno]{article}
\usepackage{theorem}
\newtheorem{thm}{THEOREM}
\newtheorem{lem}[thm]{LEMMA}

\newtheorem{pro}[thm]{PROPOSITION}
\theoremheaderfont{\scshape}

\newcommand{\ket}[1]{|#1\rangle}

\title{{\Large {\bf Quantum Random Walks in One Dimension}
}}
\author{ Norio Konno \\
}

\date{\empty }
\begin{document}
\maketitle

\par\noindent
\begin{small}
{\bf Abstract}. 
This letter treats the quantum random walk on the line determined by a $2 \times 2$ unitary matrix $U$. A combinatorial expression for the $m$th moment of the quantum random walk is presented by using 4 matrices, $P, Q, R$ and $S$ given by $U$. The dependence of the $m$th moment on $U$ and initial qubit state $\varphi$ is clarified. A new type of limit theorems for the quantum walk is given. Furthermore necessary and sufficient conditions for symmetry of distribution for the quantum walk is presented. Our results show that the behavior of quantum random walk is striking different from that of the classical ramdom walk.

\footnote[0]{
Department of Applied Mathematics, Faculty of Engineering, Yokohama National University,\\79-5 Tokiwadai, Hodogaya, Yokohama 240-8501, Japan.
}




\end{small}

\setcounter{equation}{0}

\section{Introduction}

\newcommand{\U}{\bar{U}}

\maketitle


\par
Recently various problems for quantum random walks have been widely investigated by a number of groups in connection with quantum computing. Examples include Aharonov {\it et al.},$^{(1)}$ Ambainis {\it et al.}, $^{(2)}$ Bach {\it et al.},$^{(3)}$ Childs, Farhi and Gutmann,$^{(4)}$ D\"ur {\it et al.},$^{(5)}$ Kempe,$^{(6)}$, Konno,$^{(7)}$ Konno, Namiki and Soshi,$^{(8)}$ Konno {\it et al.},$^{(9)}$ Mackay {\it et al.},$^{(10)}$ Moore and Russell,$^{(11)}$ Travaglione and Milburn,$^{(12)}$ Yamasaki, Kobayashi and Imai.$^{(13)}$ For a more general setting including quantum cellular automata, see Meyer.$^{(14)}$ A more mathematical point of view of quantum computing can be found in Brylinsky and Chen.$^{(15)}$
\par
In Ambainis {\it et al.},$^{(2)}$ they gave two general ideas for analyzing quantum random walks. One is the path integral approach, the other is the Schr\"odinger approach. In this paper, we take the path integral approach, that is, the probability amplitude of a state for the quantum random walk is given as a combinatorial sum over all possible paths leading to that state. 

\par
The quantum random walk considered here is determined by a $2 \times 2$ unitary matrix $U$ stated below. The new points of this paper is to introduce 4 matrices, $P, Q, R, S$ given by the unitary matrix $U$, to obtain a combinatorial expression for the characteristic function by using them, and to clarify a dependence of the $m$th moment of the quantum random walk on the unitary matrix $U$ and initial qubit state $\varphi$. We give a new type of limit theorems for by using our results. Moreover we present necessary and sufficient conditions for symmetry of distribution for the quantum walk. Our results show that the behavior of quantum random walk is remarkable different from that of the classical ramdom walk. \par 
The time evolution of the one-dimensional quantum random walk studied here is given by the following unitary matrix (see Nielsen and Chuang$^{(16)}$):
\begin{eqnarray*}
U=
\left[
\begin{array}{cc}
a & b \\
c & d
\end{array}
\right]
\end{eqnarray*}
\par\noindent
where $a,b,c,d \in {\bf C}$ and ${\bf C}$ is the set of complex numbers. So we have $|a|^2 + |b|^2 = |c|^2 + |d|^2 =1, \> a \overline{c} + b \overline{d}=0, \> c= - \triangle \overline{b}, \> d= \triangle \overline{a}$, where $\overline{z}$ is a complex conjugate of $z \in {\bf C}$ and $\triangle = \det U = ad - bc$ with $|\triangle|=1$. The qunatum walk is a quantum generalization of the classical random walk with an additional degree of freedom called the chirality. The chirality takes values left and right, and means the direction of the motion of the particle. The evolution of the qunatum walk is given by the following way. At each time step, if the particle has the left chirality, it moves one step to the left, and if it has the right chirality, it moves one step to the right. The unitary matrix $U$ acts on two chirality states $\ket{L}$ and $\ket{R}$:
\begin{eqnarray*}
&& \ket{L} \>\> \to \>\> a\ket{L} + c\ket{R} \\
&& \ket{R} \>\> \to \>\> b\ket{L} + d\ket{R} 
\end{eqnarray*}
where $L$ and $R$ refer to the right and left chirality state respectively. In fact, define
\begin{eqnarray*}
\ket{L} = 
\left[
\begin{array}{cc}
1 \\
0  
\end{array}
\right],
\qquad
\ket{R} = 
\left[
\begin{array}{cc}
0 \\
1  
\end{array}
\right]
\end{eqnarray*}
so we have
\begin{eqnarray*}
&& U\ket{L} = a\ket{L} + c\ket{R} \\
&& U\ket{R} = b\ket{L} + d\ket{R} 
\end{eqnarray*}
\par
More precisely, at any time $n$, the amplitude of the location of the particle is defined by a 2-vector $\in {\bf C}^2$ at each location $\{ \ldots, -2, -1, 0, 1, 2, \ldots \}$. The probability the particle is at location $k$ is given by the square of the modulus of the vector at $k$. If $\ket{\Psi_k (n)}$ define the amplitude at time $n$ at location $k$ where
\begin{eqnarray*}
\ket{\Psi_k (n)} = \left[
\begin{array}{cc}
\psi_k ^L (n) \\
\psi_k ^R (n)
\end{array}
\right]
\end{eqnarray*}
with the chirality being left (upper component) or right (lower component), then the dynamics for $\ket{\Psi_k (n)}$ in quantum walks is given by the following transformation:
\begin{eqnarray*}
\ket{\Psi_k (n+1)} = \ket{L} \langle L | U \ket{\Psi_{k+1} (n)}
+ \ket{R} \langle R | U \ket{\Psi_{k-1} (n)}
\end{eqnarray*}
The above equation can be rewritten as
\begin{eqnarray*}
\ket{\Psi_k (n+1)} = P \ket{\Psi_{k+1} (n)} + Q \ket{\Psi_{k-1} (n)}
\end{eqnarray*}
where
\begin{eqnarray*}
P= 
\left[
\begin{array}{cc}
a & b \\
0 & 0 
\end{array}
\right], 
\quad
Q=
\left[
\begin{array}{cc}
0 & 0 \\
c & d 
\end{array}
\right]
\end{eqnarray*}
We see that $U=P+Q.$ The unitarity of $U$ ensures that the amplitude always defines a probability distribution for the location.

\par
The simplest and well-studied example of a qunatum walk is the Hadamard walk whose unitary matrix $U$ is defined by 
\begin{eqnarray*}
H = 
{1 \over \sqrt{2}}
\left[
\begin{array}{cc}
1 & 1 \\
1 & -1 
\end{array}
\right] 
\end{eqnarray*}
The dynamics of this walk corresponds to that of the symmetric random walk in the classical case. In general, the following unitary matrices can also lead to symmetric walks:
\begin{eqnarray*}
U_{\eta, \phi, \psi} = {e^{i \eta} \over \sqrt{2}}
\left[
\begin{array}{cc}
e^{i(\phi + \psi)} & e^{-i(\phi - \psi)}  \\
e^{i(\phi - \psi)}  & -e^{-i(\phi + \psi)}  
\end{array}
\right] 
\end{eqnarray*}
where $\eta, \phi,$ and $\psi$ are real numbers (see, for example, pp.175-176 in Nielsen and Chuang$^{(16)}$). In particular, we see $U_{0,0,0}=H$. However symmetry of the Hadamard walk depends heavily on the initial qubit state, see Theorem 4 in this letter or Ref. 8.

\par
In the present letter, the study on the dependence of some important quantities (e.g., the $m$th moment, limit distribution) on initial qubit state is one of the essential parts, so we define the set of initial qubit states as follows:
\[
\Phi = \left\{ \varphi =
\left[
\begin{array}{cc}
\alpha \\
\beta   
\end{array}
\right]
\in 
{\bf C}^2
:
|\alpha|^2 + |\beta|^2 =1
\right\}
\]
The next section is devoted to a method given by us and some our new results.

\section{Our Method and Results}

Let $X_n ^{\varphi}$ be probability distribution of the quantum random walk starting from initial qubit state $\varphi \in \Phi$. Note that $P(X_0 ^{\varphi}=0)=1$. In order to study $P(X_n ^{\varphi} =k)$ for $n+k=$ even, we need to know the following expression. For fixed $l$ and $m$ with $l+m=n$ and $m-l=k$, 
\[
\Xi(l,m)= \sum P^{l_1}Q^{m_1}P^{l_2}Q^{m_2} \cdots P^{l_n}Q^{m_n}
\]
summed over all $l_j, m_j \ge 0$ satisfying $m_1+ \cdots +m_n=m$ and $l_1+ \cdots +l_n=l$. Note that
\[
P(X_n ^{\varphi} =k) = (\Xi(l,m) \varphi)^{\ast} (\Xi(l,m) \varphi)
\]
where $\ast$ means the adjoint operator. For example, in the case of $P(X_4 ^{\varphi} = -2)$, we have to know the expression, $\Xi (3,1) = P^3Q+P^2QP+PQP^2+QP^3$. By using $P^2 = aP, \> Q^2 = d Q$, the above equation becomes $\Xi (3,1) = a^2 PQ + a PQP+ a PQP+ d^2 QP.$ In our treatment of quantum walks, as well as the matrices $P$ and $Q$, it is convenient to introduce
\[
R=
\left[
\begin{array}{cc}
c & d \\
0 & 0 
\end{array}
\right], 
\>\>
S=
\left[
\begin{array}{cc}
0 & 0 \\
a & b 
\end{array}
\right]
\]
The next table of products of $P,Q,R,$ and $S$ is very useful in computing some quantities:
\par
\
\par
\begin{center}
\begin{tabular}{c|cccc}
  & $P$ & $Q$ & $R$ & $S$  \\ \hline
$P$ & $aP$ & $bR$ & $aR$ & $bP$  \\ 
$Q$ & $cS$ & $dQ$& $cQ$ & $dS$ \\
$R$ & $cP$ & $dR$& $cR$ & $dP$ \\
$S$ & $aS$ & $bQ$ & $aQ$ & $bS$ 
\end{tabular}
\end{center}
where $PQ=bR$. We should remark that $P,Q,R,$ and $S$ form an orthonormal basis of the vector space of complex $2 \times 2$ matrices $M_2 ({\bf C})$ with respect to the trace inner product $\langle A|B \rangle = \>$ tr $(A^{\ast}B)$. So $\Xi (l,m)$ has the following form:
\[
\Xi (l,m) = p_n (l,m) P + q_n (l,m) Q + r_n (l,m) R + s_n (l,m) S
\]
Next problem is to obtain explicit forms of $p_n (l,m), q_n (l,m), r_n (l,m)$ and $s_n (l,m)$. For example, we have $\Xi (3,1) = 2abc P + a^2b R + a^2c S,$ therefore, $p_4 (3,1)=2abc, \> q_4 (3,1)=0, \> r_4 (3,1)=a^2b, \> s_4 (3,1)=a^2c .$
 In $abcd=0$ case, the argument is much easier. So from now on we focus only on $abcd \not= 0$ case. In this case, the next key lemma is obtained by a combinatorial method.
\par
\
\par\noindent
\begin{lem}
\label{lem:lem1} We consider quantum random walks in one dimension with $abcd \not= 0$. Suppose that $l,m \ge 0$ with $l+m=n$, then we have 
\par\noindent
\hbox{(i)} for $l \wedge m (= \min \{l,m \}) \ge 1$,  
\begin{eqnarray*}
\Xi (l,m) 
= a^l \overline{a}^m \triangle^m 
\sum_{\gamma =1} ^{l \wedge m} 
\left(-{|b|^2 \over |a|^2} \right)^{\gamma}
{l-1 \choose \gamma- 1} 
{m-1 \choose \gamma- 1} 
\times 
\Biggl[ {l- \gamma \over a \gamma } P + {m - \gamma \over \triangle \overline{a} \gamma} Q  - {1 \over \triangle \overline{b}} R + {1 \over b} S 
\Biggr]
\end{eqnarray*}
\par\noindent
\hbox{(ii)} for $l (=n) \ge 1, m = 0$,  
\[
\Xi (l,0) = a^{l-1} P
\]
\par\noindent
\hbox{(iii)} for $l = 0, m (=n) \ge 1$,  
\[
\Xi (0,m) = \triangle^{m-1} \overline{a}^{m-1} Q
\]
\end{lem}
By this lemma, the characteristic function of $X_n ^{\varphi}$ for $abcd \not=0$ case is obtained. Moreover, the $m$th moment of $X_n ^{\varphi}$ can be also derived from the characteristic function in the standard fashion. Here we give only the result of the $m$th moment. Result on the characteristic function and proofs of our results mentioned here appear in our mathematical paper.$^{(7)}$
\par
\
\par\noindent
\begin{pro}
\label{thm:pro2} 
We consider quantum random walks with $abcd \not= 0$.
\par\noindent
\hbox{(i)} When $m$ is odd, we have
\begin{eqnarray*}
E((X_n ^{\varphi}) ^m) 
&=& 
|a|^{2(n-1)}
\biggl[ - n^m \left\{ \left(|a|^2 - |b|^2 \right)  
\left(|\alpha|^2 - |\beta|^2 \right) 
+ 2 (a \alpha \overline{b \beta} + \overline{a \alpha} b \beta ) 
\right\}
\biggr] \\
&& + \sum_{k=1}^{\left[{n-1 \over 2}\right]}
\sum_{\gamma =1} ^{k} \sum_{\delta =1} ^{k}
\left(-{|b|^2 \over |a|^2} \right)^{\gamma + \delta} 
{k-1 \choose \gamma- 1} 
{k-1 \choose \delta- 1} 
{n-k-1 \choose \gamma- 1} 
{n-k-1 \choose \delta- 1} \\
&& 
\times {(n-2k)^{m+1} \over  \gamma \delta} 
\Biggl[ 
 - \{ n( |a|^2 - |b|^2 ) + \gamma + \delta \} (|\alpha|^2 - |\beta|^2 ) \\
&&
\qquad \qquad \qquad \qquad \qquad \qquad 
+ \left( {\gamma +\delta \over |b|^2} - 2n \right) 
( a \alpha \overline{b \beta} + \overline{a \alpha} b \beta )
\Biggr]
\end{eqnarray*}
\par\noindent
\hbox{(ii)} When $m$ is even, we have
\begin{eqnarray*}
E((X_n^{\varphi}) ^m) 
&=&  |a|^{2(n-1)} 
\Biggl[ 
n^m \\
&&
+
\sum_{k=1}^{\left[{n-1 \over 2}\right]}
\sum_{\gamma =1} ^{k} \sum_{\delta =1} ^{k}
\left(-{|b|^2 \over |a|^2} \right)^{\gamma + \delta} 
{k-1 \choose \gamma- 1} 
{k-1 \choose \delta- 1} 
{n-k-1 \choose \gamma- 1} 
{n-k-1 \choose \delta- 1} \\
&& 
\times {(n-2k)^{m} \over  \gamma \delta} 
\biggl\{
(n-k)^2 + k^2 - n (\gamma + \delta) + {2 \gamma \delta \over |b|^2} 
\biggr\}
\Biggr]
\end{eqnarray*}
\end{pro}
\par
\
\par\noindent
It should be noted that for any case, when $m$ is even, $E((X_n ^{\varphi})^m)$ is independent of initial qubit state $\varphi$. Moreover we have the following new type of limit theorems:
\par
\
\par\noindent
\begin{thm}
\label{thm:thm3} We assume $abcd \not= 0$. If $n \to \infty$, then 
\[
{X_n ^{\varphi} \over n} \quad \Rightarrow \quad Z^{\varphi}
\]
where $Z^{\varphi}$ has a density 
\[
f(x; {}^t[\alpha, \beta])
= { \sqrt{1 - |a|^2} \over \pi (1 - x^2) \sqrt{|a|^2 - x^2}} 
 \left\{ 1- \left( |\alpha|^2 - |\beta|^2 + {a \alpha \overline{b \beta} + \overline{a \alpha} b \beta \over |a|^2 } \right) x \right\} 
\]
for $x \in (- |a|, |a|)$ with 
\begin{eqnarray*}
&& E(Z^{\varphi}) 
=
- \left( |\alpha|^2 - |\beta|^2 + { a \alpha \overline{b \beta} 
+ \overline{a \alpha} b \beta \over |a|^2 } \right) 
\times (1 - \sqrt{1 - |a|^2}) 
\\
&& E ((Z^{\varphi})^2) = 1 - \sqrt{1 - |a|^2}
\end{eqnarray*}
and $ Y_n \Rightarrow Y$ means that $Y_n$ converges in distribution to a limit $Y$. 
\end{thm}
\par
\
\par\noindent
We remark that standard deviation of $Z^{\varphi}$ is not independent of initial qubit state $\varphi ={}^t[\alpha, \beta]$. The above limit theorem suggests the following result on symmetry of distribution for quantum random walks. This is a generalization of the result give by Konno, Namiki and Soshi$^{(8)}$ for the Hadamard walk. Define
\begin{eqnarray*}
\Phi_s &=&  \{ \varphi \in 
\Phi : \> 
P(X_n ^{\varphi}=k) = P(X_n ^{\varphi}=-k) \>\> 
\hbox{for any} \> n \in {\bf Z}_+ \> \hbox{and} \> k \in {\bf Z}
\} 
\\
\Phi_0 &=& \left\{ \varphi \in 
\Phi : \> 
E(X_n ^{\varphi})=0 \>\> \hbox{for any} \> n \in {\bf Z}_+
\right\}
\\
\Phi_{\bot} &=& \left\{ \varphi = {}^t[\alpha, \beta] \in 
\Phi :
|\alpha|= |\beta|, \> a \alpha \overline{b \beta} + \overline{a \alpha} b \beta =0 
\right\} 
\end{eqnarray*}
and ${\bf Z}$ (resp. ${\bf Z}_+$) is the set of (resp. non-negative) integers.  For $\varphi \in \Phi_s$, the probability distribution of $X_n ^{\varphi}$ is symmetric for any $n \in {\bf Z}_+$. Using the explicit form of $E(X_n ^{\varphi})$ given by Proposition 2 (i) ($m=1$ case), we have 
\par
\
\par\noindent
\begin{thm}
\label{thm:thm4} We assume $abcd \not= 0$. Then we have
\[
\Phi_{s} = \Phi_0 = \Phi_{\bot}
\]
\end{thm}

From now on we give an outline of our proof of Theorem 3 (for more details, see Konno$^{(7)}$). To do so, we introduce the Jacobi polynomial $P^{\nu, \mu} _n (x)$, where $P^{\nu, \mu} _n (x)$ is orthogonal on $[-1,1]$ with respect to $(1-x)^{\nu}(1+x)^{\mu}$ with $\nu, \mu > -1$. 

First we obtain the next asymptotics of characteristic function $E(e^{i \xi X^{\varphi}_n/n})$ by using the Jacobi polynomial: if $n \to \infty$ with $k/n=x \in (-(1-|a|)/2, (1+|a|)/2)$, then
\begin{eqnarray*}
&& E(e^{i \xi X_n ^{\varphi}/n}) 
\sim 
\sum_{k=1}^{\left[{n-1 \over 2}\right]} |a|^{2n - 4k -2}|b|^4 \\
&& \times 
\biggl[ 
\left\{ {2x^2-2x+1 \over x^2} (P^{1,n-2k} _{k-1})^2 - {2 \over x} P^{1,n-2k} _{k-1} P^{0,n-2k} _{k-1}+
{2 \over |b|^2} (P^{0,n-2k} _{k-1})^2 \right\} \cos ((1-2x) \xi) \\
&& 
+ \left( {1-2x \over x} \right) 
\biggl\{ - {1 \over x} 
\{ 
(|a|^2 - |b|^2) 
(|\alpha|^2 - |\beta|^2) + 2 (a \alpha \overline{b \beta} + \overline{a \alpha} b \beta ) \} (P^{1,n-2k} _{k-1})^2 \\
&&
-2 \left( |\alpha|^2 - |\beta|^2 - 
{ a \alpha \overline{b \beta} + \overline{a \alpha} b \beta \over |b|^2 } 
\right) P^{0,n-2k} _{k-1} P^{1,n-2k} _{k-1} \biggl\} 
i \sin ((1-2x) \xi)
\biggr]
\end{eqnarray*}
where $f(n) \sim g(n) $ means $f(n)/g(n) \to 1 \> (n \to \infty)$, and $P^{i,n-2k} _{k-1} = P^{i,n-2k} _{k-1}(2|a|^2-1) \> (i=0,1)$.
\par
Next we prepare the following asymptotic results for the Jacobi polynomial $P^{\alpha + an, \beta +bn} _n (x)$ derived by Chen and Ismail,$^{(17)}$: if $n \to \infty$ with $k/n=x \in (-(1-|a|)/2, (1+|a|)/2)$, then
\begin{eqnarray*}
&& P^{0,n-2k} _{k-1} \sim  
{ 2 |a|^{2k-n} \over \sqrt{\pi n \sqrt{- \Delta}} } \cos (An+B)  \\
&& P^{1,n-2k} _{k-1} \sim  
{2 |a|^{2k-n} \over \sqrt{\pi n \sqrt{- \Delta}}} \sqrt{{x \over (1-x)(1-|a|^2)}} \cos (An+B+ \theta) 
\end{eqnarray*}
where $\Delta = (1-|a|^2)(4x^2-4x+1-|a|^2)$, $A$ and $B$ are some constants (which are independent of $n$), and $\theta \in [0, \pi/2]$ is determined by $\cos \theta = \sqrt{(1-|a|^2)/4x(1-x)}$. 

\par
Combining these results, Theorem 3 is obtained. 
\par
Now we compare our analytical results (Theorem 3) with the numerical ones$^{(10,12)}$ for the the Haramard walk. In this case, Theorem 3 implies that if $ -\sqrt{2}/2 <a <b <\sqrt{2}/2$, then as $n \to \infty$, 
\begin{eqnarray*}
P(a \le X^{\varphi} _{n}/n \le b)  \to  \int^b _a {  1-  (|\alpha|^2 - |\beta|^2 + \alpha \overline{\beta} + \overline{\alpha} \beta) x \over \pi (1-x^2) \sqrt{1-2x^2}} dx
\end{eqnarray*}
for any initial qubit state $\varphi = {}^t[\alpha, \beta]$. For the classical symmetric random walk $Y^o _n$ starting from the origin, the well-known central limit theorem implies that if $ - \infty <a <b < \infty $, then as $n \to \infty$, 
\begin{eqnarray*}
P(a \le Y^o _{n}/ \sqrt{n} \le b)  \to \int^b _a {e^{-x^2/2} \over \sqrt{2 \pi}} dx\end{eqnarray*}
This result is often called the de Moivre-Laplace theorem. When we take $\varphi = {}^t[1/\sqrt{2},i/\sqrt{2}]$ (symmetric case), then we have the following quantum version of the de Moivre-Laplace theorem: if $ -\sqrt{2}/2 <a <b <\sqrt{2}/2$, then as $n \to \infty$, 
\begin{eqnarray*}
P(a \le X^{\varphi} _{n}/n \le b) \to \int^b _a {1 \over \pi (1-x^2) \sqrt{1-2x^2}} dx
\end{eqnarray*}
So there is a remarkable difference between the quantum random walk $X^{\varphi} _n$ and the classical one $Y^o _n$ even in a symmetric case for $\varphi = {}^t[1/\sqrt{2},i/\sqrt{2}]$. 
\par
Noting that $E(X^{\varphi} _n)=0 \> (n \ge 0)$ for any $\varphi \in \Phi_{\bot}$, we have
\begin{eqnarray*}
sd(X_n ^{\varphi})/n \to  \sqrt{(2 - \sqrt{2})/2} = 0.54119 \ldots 
\end{eqnarray*}
where $sd(X)$ is the standard deviation of $X$. This rigorous result reveals that numerical simulation result 3/5 = 0.6 given by Travaglione and Milburn$^{(12)}$ is not so accurate.
\par
As in a similar way, when we take $\varphi = {}^t[0,e^{i \theta}]$ where $\theta \in [0, 2\pi)$ (asymmetric case), we see that if $ -\sqrt{2}/2 <a <b <\sqrt{2}/2$, then as $n \to \infty$, 
\begin{eqnarray*}
P(a \le X^{\varphi} _{n}/n \le b)  \to  \int^b _a {1 \over \pi (1-x) \sqrt{1-2x^2}} dx
\end{eqnarray*}
So we have
\begin{eqnarray*}
E(X_n ^{\varphi})/n  \to  (2 - \sqrt{2})/2 = 0.29289 \ldots , \quad 
sd(X_n ^{\varphi})/n  \to \sqrt{(\sqrt{2} -1)/2} = 0.45508 \ldots 
\end{eqnarray*}
When $\varphi = {}^t[0,1]$ ($\theta =0$), Ambainis {\it et al.}$^{(2)}$ gave the same result. In the paper, they took two approaches, that is, the Schr\"odinger approach and the path integral approach. However their result comes mainly from the Schr\"odinger approach by using a Fourier analysis. The details on the derivation based on the path integral approach is not so clear compared with Konno$^{(7)}$. 

\par
In another asymmetric case $\varphi = {}^t[e^{i \theta},0]$ where $\theta \in [0, 2\pi)$, a similar argument implies that if $ -\sqrt{2}/2 <a <b <\sqrt{2}/2$, then as $n \to \infty$, 
\begin{eqnarray*}
P(a \le X^{\varphi} _{n}/n \le b)  \to  \int^b _a {1 \over \pi (1+x) \sqrt{1-2x^2}} dx
\end{eqnarray*}
Note that $f(-x ; {}^t[e^{i \theta},0])= f(x; {}^t[0, e^{i \theta}])$ for any $x \in (-\sqrt{2}/2, \sqrt{2}/2)$. Therefore concerning the $m$th moment of the limit distribution, we have the same result as in the previous case $\varphi = {}^t[0, e^{i \theta}]$. So the standard deviation of the limit distribution $Z^{\varphi}$ is given by $\sqrt{(\sqrt{2} -1)/2} = 0.45508 \ldots.$ Simulation result 0.4544 $\pm$ 0.0012 in Mackay {\it et al.}$^{(10)}$ (their case is $\theta =0$) is consistent with our rigorous result.

\par
\
\par\noindent
{\bf Acknowledgments.}

This work is partially financed by the Grant-in-Aid for Scientific Research (B) (No.12440024) of Japan Society of the Promotion of Science. I would like to thank Takao Namiki, Takahiro Soshi, Hideki Tanemura, and Makoto Katori for useful discussions.


\par
\
\par\noindent

\begin{small}

\bibliographystyle{plain}

\end{small}

\end{document}